\documentclass[12pt]{article}
\usepackage{pdproc,epsfig} 

\begin{document}

\renewcommand{\thefootnote}{\alph{footnote}}

\hfill{IFIC/03-28}
  
\title{RELIC NEUTRINOS: NEUTRINO\\  
PROPERTIES FROM COSMOLOGY
\footnote{Invited talk at the X Int.\ Workshop on Neutrino
Telescopes, Venice, March 11-14, 2003.}}

\author{SERGIO PASTOR}

\address{ Instituto de F\'{\i}sica Corpuscular (CSIC-Universitat de
Val\`encia)\\ Ed.\ Institutos de Investigaci\'on, Apdo.\ 22085,
E-46071 Valencia, Spain\\
 {\rm E-mail: Sergio.Pastor@ific.uv.es}}

\abstract{After a short introduction on the predicted cosmic neutrino
background in the universe, we review some of the cosmological bounds
related to neutrinos. In particular we show how the recent data on the
anisotropies of the cosmic microwave background radiation from the
WMAP satellite, combined with other experimental results, can
constrain the radiation content of the universal energy density and
the sum of neutrino masses.}
   
\normalsize\baselineskip=15pt

\section{Introduction}

Neutrinos are the second most abundant particles in the universe, with
a number density comparable to that of relic photons. Therefore, the
relic neutrino sea influences various cosmological situations, playing
an important role that has been discussed in many papers, both to
solve some cosmological problems and to put bounds on non-standard
neutrino properties. These cosmological limits are alternative to the
data from terrestrial neutrino experiments, in some cases the only
available.

Here we review recent bounds from cosmology on the number of neutrinos
(content of relativistic particles) and on the neutrino mass, in
particular those that appeared after the release of the first year
data of the Wilkinson Microwave Anisotropy Probe (WMAP)
\cite{Bennett:2003bz}. There exist many other topics not discussed
here, such as the role of massive neutrinos in leptogenesis, that can
be found in the extensive review \cite{Dolgov:2002wy} and in the more
recent ones \cite{Raffelt:2003nc,Sarkar:2003ch}.

\section{Standard relic neutrinos}

In the early universe, at temperatures above MeV, the neutrinos were
in thermal equilibrium through the standard weak interactions with
other particles. Thus the distribution of neutrino momenta was a
Fermi-Dirac one,
\begin{equation}
f_{\nu_\alpha}(p)=\left[\exp\left (\frac{p-\mu_\alpha}{T}\right
)+1\right]^{-1}
\label{feq}
\end{equation}
where the energy $E\simeq p$ for neutrino masses much smaller than
MeV.  The $\mu_\alpha$'s are the chemical potentials of the different
neutrino flavours, which could have been non-zero if an asymmetry
between the number of neutrinos and antineutrinos was previously
created. We neglect this possibility in this section, but it will be
considered later.

As the universe cools, the weak interaction rate $\Gamma_\nu$ falls
below the expansion rate $H$ and the neutrinos decouple from the rest
of the plasma. A rough but quite good estimate can be found equating
the average value of $\Gamma_\nu=\langle\sigma_\nu n_\nu\rangle$ and
$H=\sqrt{8\pi\rho/3M_P^2}$, which gives $T_{\rm dec}\simeq {\cal
O}$(MeV).  A better calculation gives $T_{\rm dec}\simeq 2-3$ MeV,
where the higher (lower) decoupling temperature corresponds to
$\nu_{\mu,\tau}$ ($\nu_e$). After decoupling, the collisionless
neutrinos expand freely, keeping their phase-space density as in
Eq.~(\ref{feq}), since both momentum and temperature redshift
identically. Finally, at temperatures of order the electron mass, the
electron--positron pairs annihilate and transfer their entropy into
photons but not into the decoupled neutrinos, causing the well-known
difference between the temperatures of relic photons and relic
neutrinos, $T_\gamma/T_\nu = \left( 11/4 \right)^{1/3}$ (see e.g.\
\cite{kt}).

Note that the relic neutrino phase-space density in Eq.~(\ref{feq}) is
that of relativistic particles (or radiation) in equilibrium, valid
also for massive neutrinos when $m_\nu \gg T_\nu$. It is easy to
calculate the number and energy densities of relic neutrinos at any
temperature $T=T_\gamma$ below MeV. The number density per flavour is
fixed by the value of the temperature, 
\begin{equation}
n_{\nu} = \frac{3}{11}\;n_\gamma = \frac{6\zeta(3)}{11\pi^2}\;T^3
\label{nunumber}
\end{equation}
while the energy density for massive neutrinos should be in principle
calculated numerically, with two well-defined analytical limits,
$$
\rho_\nu (m_\nu \ll T_\nu)= 
\frac{7\pi^2}{120}
\left(\frac{4}{11}\right)^{4/3}\;T^4
$$
\begin{equation}
\rho_\nu (m_\nu \gg T_\nu)=  m_\nu n_\nu
\label{nurho}
\end{equation}
Thus we see that the contribution of massive neutrinos to the energy
density in the non-relativistic limit is a function of the mass (or
the sum of masses if all neutrino states have $m_i \gg T_\nu$).

Relic neutrinos contribute to the relativistic energy density of the
universe while $m_\nu \ll T_\nu$, fixing the expansion rate during the
production of primordial abundances of light elements (Big Bang
Nucleosynthesis or BBN). BBN is the last period of the universe
sensitive to neutrino flavour, since electron neutrinos and
antineutrinos play a direct role in the freezing of the
neutron-to-proton ratio. Later cosmological stages, such as the
formation of the angular fluctuations of the cosmic microwave
background (CMB) and the formation of large-scale structures (LSS),
are also sensitive to relativistic relic neutrinos but blind to
neutrino flavour. In particular the small-scale power of the cosmic
density fluctuations is modified by neutrino masses of order eV and
smaller.

\section{Number of relativistic neutrino species}

The energy density of the universe stored in relativistic species,
$\rho_r$, is customarily given in terms of the so-called {\it
effective number of relativistic neutrino species} $N_{\rm eff}$ (see
\cite{Dolgov:2002wy} for a review and references), through the
relation
\begin{equation}
\rho_{\rm r} = \rho_\gamma + \rho_\nu + \rho_x =
\left[ 1 + \frac{7}{8} \left( \frac{4}{11}
\right)^{4/3} \, N_{\rm eff} \right] \, \rho_\gamma \,\,,
\label{neff}
\end{equation}
where $\rho_\gamma$ is the energy density of photons, whose value
today is known from the measurement of the CMB temperature. Eq.\
(\ref{neff}) can be also written as
\begin{equation}
N_{\rm eff} \equiv \left( \frac{\rho_r -\rho_\gamma}{\rho^0_\nu}\right)
\left(\frac{\rho^0_\gamma}{\rho_\gamma} \right)\, ,
\label{neff-def}
\end{equation}
where $\rho_\nu^0$ denotes the energy density of a single species of
massless neutrino with an equilibrium Fermi-Dirac distribution with
zero chemical potential, and $\rho^0_\gamma$ is the photon energy
density in the approximation of instantaneous neutrino decoupling. The
normalization of $N_{\rm eff}$ is such that it gives $N_{\rm eff} = 3$
in the standard case of three flavours of massless neutrinos, again in
the limit of instantaneous decoupling. In principle $N_{\rm eff}$
includes, in addition to the standard neutrinos, a potential
contribution $\rho_x$ from other relativistic relics such as majorons
or sterile neutrinos.

It turns out that even in the standard case of three neutrino flavours
the effective number of relativistic neutrino species is not exactly
3. The decoupling of neutrinos from the rest of the primordial plasma
occurs not far from the $e^\pm$ annihilations, and accurate
calculations \cite{Dolgov:1997mb} have shown that neutrinos are still
slightly interacting, thus sharing a small part of the entropy
release. This causes a momentum dependent distortion in the neutrino
spectra from the equilibrium Fermi--Dirac behavior and a slightly
smaller $T/T_\nu$ ratio. Both effects lead to a value of $N_{\rm eff}
= 3.034$.  A further, though smaller, effect on $T/T_\nu$ is induced
by finite temperature Quantum Electrodynamics (QED) corrections to the
electromagnetic plasma. A recent combined study \cite{Mangano:2001iu}
of the incomplete neutrino decoupling and QED corrections concluded
that the total effect corresponds to $N_{\rm eff} = 3.0395 \simeq
3.04$ . Therefore we define the extra energy density in radiation form
as
\begin{equation}
\Delta N_{\rm eff} \equiv N_{\rm eff}-3.04 \, \, .
\label{delta-neff}
\end{equation}
The standard value of $N_{\rm eff}$ corresponds to the case of
massless or very light neutrinos, i.e.\ those with masses much smaller
than 1 eV. More massive neutrinos affect the late evolution of the
universe in a way that can not be parametrized with a $\Delta N_{\rm
eff}$, as described in the next section.

The value of $\Delta N_{\rm eff}$ is constrained at the BBN epoch from
the comparison of theoretical predictions and experimental data on the
primordial abundances of light elements, which also depend on the
baryon-to-photon ratio $\eta=n_B/n_\gamma$ (or baryon density). The
main effect of a non-zero $\Delta N_{\rm eff}$ is to modify the Hubble
expansion rate through its contribution to the total energy density
$H=\sqrt{8\pi\rho/3M_P^2}$. This in turn changes the freezing
temperature of the neutron-to-proton ratio, therefore producing a
different abundance of $^4$He. Typically, the BBN bounds are of order
$\Delta N_{\rm eff} < 0.4-1$
\cite{Lisi:1999ng,Esposito:2000hh,Kneller:2001cd,Cyburt:2001pq}.

\begin{figure}[t]
\hspace{0.225\textwidth}
\epsfig{figure=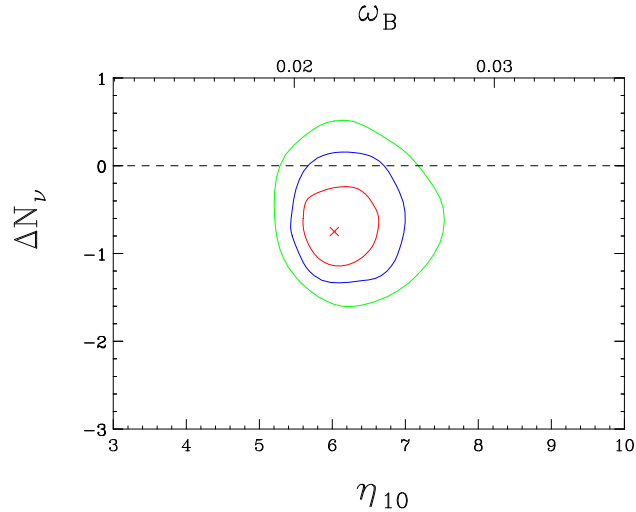,width=0.55\textwidth}
\caption{The $1\sigma$, $2\sigma$ and $3\sigma$ contours in the
$\eta_{10}-\Delta N_{\rm eff}$ plane from a combination of BBN and
WMAP data. Here $\omega_B=\Omega_Bh^2$ and
$\eta_{10}=10^{10}\eta_B\simeq 274\;\omega_B$. Figure from ref.\
\protect\cite{Barger:2003zg}.
\label{cmb+bbn} }
\end{figure}

The BBN bounds on $\Delta N_{\rm eff}$ have been recently reanalyzed
using as input the value of the baryon density derived from WMAP data
\cite{Spergel:2003cb} $\eta_{\rm CMB}=6.14\pm 0.25$. Hannestad
\cite{Hannestad:2003xv} finds the constraint $N_{\rm
eff}=2.6^{+0.4}_{-0.3}$ (95\% CL), in agreement with the results of
refs.\ \cite{Cyburt:2003fe,Barger:2003zg}. The allowed regions are
shown in Fig.\ \ref{cmb+bbn}, taken from ref.\ \cite{Barger:2003zg}.
The bounds suggest that the preferred value of $\Delta N_{\rm eff}$ is
below zero, i.e.\ inconsistent with the standard case. However, it is
probably premature to consider this a robust indication of
non-standard physics, because the observational data on primordial
$^4$He is still dominated by systematics. It is also interesting 
that, since the observed primordial abundance of deuterium agrees
well with the value predicted from $\eta_{\rm CMB}$, one can now use D
to obtain a limit on $N_{\rm eff}$. The authors of
\cite{Cyburt:2003fe} found, for an averaged D abundance, the limits
$1.22<N_{\rm eff}<5.25$ (95\% CL), in agreement with the standard
prediction. In the future, improved bounds can be obtained for more
precise values of $\eta_{\rm CMB}$ and the primordial abundance of D.

It is important to note that the BBN bounds on $\Delta N_{\rm eff}$
change if the $\nu_e$ or $\bar{\nu}_e$ spectrum show deviations from
the standard Fermi-Dirac form with $\mu_{\nu_e}=0$. For details, see for
instance the recent discussion in \cite{DiBari:2003fg}.

Independent bounds on the radiation content of the universe at a later
epoch can be extracted from the analysis of the power spectrum of CMB
anisotropies.  An enhanced contribution of relativistic particles
delays the epoch of matter-radiation equality, which in turn increases
the early integrated Sachs-Wolfe effect. Basically this leads to more
power around the scale of the first CMB peak. Previous analyses found
weak bounds on $\Delta N_{\rm eff}$
\cite{Jungman:1995bz,Esposito:2000sv,Hannestad:2001hn,Hansen:2001hi},
that can be significantly improved by adding priors on the age of the
universe or by including supernovae and LSS data \cite{Hu:1998tk}.
Before the WMAP data, from a combination of CMB and LSS data from the
PSCz redshift survey, ref.\ \cite{Hannestad:2001hn} found the allowed
range $N_{\rm eff}=6^{+8}_{-4.5}$ (95\% CL). 

\begin{figure}[t]
\hspace{0.175\textwidth}
\epsfig{figure=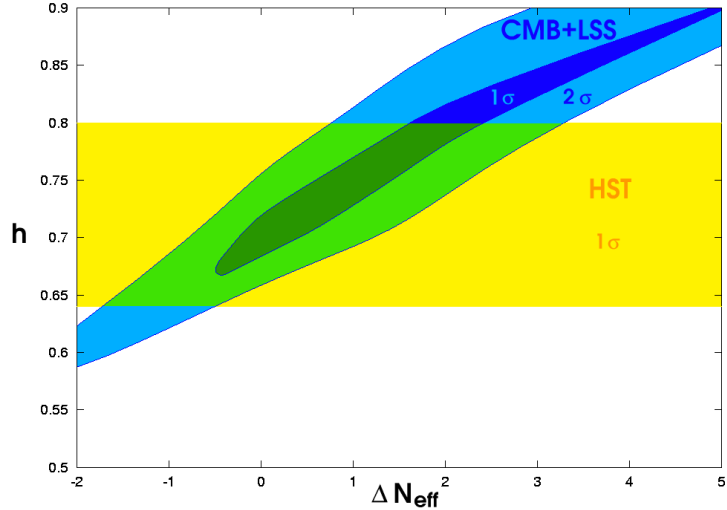,width=0.65\textwidth}
\caption{\label{deltaN_h} The two-dimensional confidence limits on
($\Delta N_{\rm eff}$, $h$), based on CMB and LSS data, at the
1-$\sigma$ (dark blue) and 2-$\sigma$ (light blue) levels. The
superimposed yellow stripe shows the HST result. Figure from ref.\
\protect\cite{Crotty:2003th}.}
\end{figure}

The inclusion of the WMAP data has significantly tightened the bounds,
as shown in refs.\ \cite{Crotty:2003th,Hannestad:2003xv}. Assuming a
minimal set of cosmological parameters and a flat universe, the limit
is now $\Delta N_{\rm eff}=0.5^{+3.3}_{-2.1}$ (95\% CL), a value found
using data from WMAP and other CMB experiments, as well as LSS data
derived from the 2dF Galaxy Redshift Survey (2dFGRS). In particular, a
parameter degeneracy exists between $\Delta N_{\rm eff}$ and the
present value of the Hubble constant $h\equiv H_0/(100$ km s$^{-1}$
Mpc$^{-1})$, which is partially removed including the results of the
Hubble Space Telescope Key Project \cite{HST} (HST), $h=0.72\pm 0.08$
at $1\sigma$. This degeneracy is shown in Fig.\ \ref{deltaN_h}, taken
from ref.\ \cite{Crotty:2003th} (see also \cite{Hannestad:2001hn}).
In the case of a non-flat universe \cite{Pierpaoli:2003kw}, the
corresponding limit is modified to $\Delta N_{\rm
eff}=1.1^{+2.0}_{-1.9}$.

Although the recent CMB+LSS bounds on $\Delta N_{\rm eff}$ are not as
restrictive as those from BBN, they will be significantly improved in
the future with the next measurements of the CMB temperature and
polarization spectra by WMAP (and those of PLANCK in the near future),
as shown in several forecast analyses
\cite{Lopez:1998aq,Kinney:1999pd,Bowen:2001in}. An analysis for models
without dark energy contribution \cite{Lopez:1998aq} estimated that it
would be possible to measure $\Delta N_{\rm eff} \sim 0.04$, the
standard value calculated with non-instantaneous neutrino
decoupling. However, this is probably too optimistic, and a more
recent estimate in ref.\ \cite{Bowen:2001in} gives $\Delta N_{\rm eff}
\sim 0.2$ with the inclusion of PLANCK data.

Many extensions of the Standard Model of particle physics predict
additional relativistic degrees of freedom that will contribute to
$\Delta N_{\rm eff}$. There exist models with 4 neutrinos which
include an additional sterile neutrino in order to explain the third
experimental indication of neutrino oscillations (the LSND results).
It was shown in many studies (see for instance
\cite{DiBari:2001ua,Abazajian:2002bj}) that all four neutrino models,
both of 2+2 and 3+1 type, lead to a full thermalization of the sterile
neutrino before BBN, and thus to $\Delta N_{\rm eff}\simeq 1$,
a value disfavoured in the standard minimal model of BBN.  Moreover, in
these models there exists at least one neutrino state with mass of
order 1 eV, and some of the regions of parameters are also disfavoured
by the cosmological upper limit on $\sum m_\nu$, as shown in the next
section.

It is also possible that the relativistic degrees of freedom at the
BBN and CMB epochs differ, for instance because of particle decays
which increase the photon temperature relative to the neutrino one
\cite{Kaplinghat:2000jj}. In some situations $\Delta N_{\rm eff}$ can
be effectively negative at BBN, such as the case of a distortion in
the $\nu_e$ or $\bar{\nu}_e$ spectra
\cite{Dolgov:1998st,Hansen:2000td}, or a very low reheating scenario
\cite{Giudice:2000ex}. 

A non-standard case that has been considered many times in the past is
the existence of relic neutrino asymmetries, namely when the number of
neutrinos and antineutrinos of the same flavour is significantly
different. These so-called degenerate neutrinos are described by a
dimensionless chemical potential $\xi_\alpha=\mu_{\nu_\alpha}/T$, and
it has been shown that the neutrino energy density always increases
for any value $\xi_\alpha\neq 0$
\begin{equation}
\Delta N_{\rm eff} = \frac{15}{7} \sum_\alpha \left [ 2 \left
(\frac{\xi_\alpha}{\pi}\right )^2 + \left
(\frac{\xi_\alpha}{\pi}\right )^4 \right ]
\label{delta-neff-xi}
\end{equation}
Interestingly, some combinations of pairs $(\xi_e,\xi_{\mu,\tau})$
could still produce the primordial abundances of light elements for a
larger baryon asymmetry, in the so-called degenerate BBN scenario
\cite{Kang}. At the same time the weaker CMB bounds on $\xi_\nu$ are
flavour blind \cite{Kinney:1999pd,paper1}. However, it was recently
shown that for neutrino oscillation parameters in the regions favoured
by atmospheric and solar neutrino data flavour equilibrium between all
active neutrino species is established well before the BBN epoch
\cite{Dolgov:2002ab,Wong:2002fa,Abazajian:2002qx}. Thus the stringent
BBN bounds on $\xi_e$ apply to all flavours, so that the contribution
of a potential relic neutrino asymmetry to $\Delta N_{\rm eff}$ is
limited to very low values. In addition, these bounds fix the value of
the neutrino number density down to 1\%, an important input to find an
upper limit on $m_\nu$.

\section{Massive relic neutrinos}

The role of neutrinos as the dark matter (DM) particles has been
widely discussed since the early 1970s. There are two facts in favour
of neutrinos as DM: they definitely exist and an eV neutrino mass
produces a contribution of order unity to the present energy density
of the universe. This can be easily showed calculating the neutrino
fraction of the critical density, $\Omega_\nu=\rho_\nu/\rho_c$. For
values of neutrino masses much larger than the present cosmic
temperature ($T_\gamma\sim T_\nu\approx 10^{-4}$ eV), one finds
\begin{equation}
\Omega_\nu h^2 = \frac{\sum_{i} m_{\nu_i}}{92.5~{\rm eV}}
\label{Omeganu}
\end{equation}
{}From this relation, one gets the traditional upper limit on neutrino
masses, of the order some tens of eV, using the very conservative
bound $\Omega_\nu<1$ (see \cite{Dolgov:2002wy,Sarkar:2003ch} for a
list of references). As discussed below, much tighter constraints are
found from LSS observations.

The evolution of cosmological perturbation is sensitive to the energy
density of massive neutrinos. The neutrino background erases the
density contrasts on wavelengths smaller than a mass--dependent
free--streaming scale. Thus massive neutrinos produce a damping of the
total density fluctuations on small scales, characteristic of the
so-called hot dark matter (HDM) particles. In a universe dominated by
HDM, the formation of structures follows a top-down scenario, where
large objects such as superclusters of galaxies form first, while
smaller structures like clusters and galaxies form via a fragmentation
process.

In the late 1970s and early 1980s, neutrinos looked like a good DM
candidate. But by the mid-1980s it was realized that HDM models were
not able to explain some of the observed structures, among others the
fact that galaxies seem older than clusters. The attention then turned
to cold dark matter (CDM) candidates, i.e.\ particles which were
non-relativistic at the epoch when the universe became
matter-dominated, which provide excellent agreement with
observations. Still in the mid-1990s it appeared that a small mixture
of HDM in a universe dominated by CDM fitted better the observational
data than a pure CDM model. However, within the presently favoured
$\Lambda$CDM model (dominated by dark energy) there is no need for a
significant contribution of HDM\footnote{For a historical review of
neutrinos as DM, see for instance \cite{Primack:2001ib}.}.

\begin{figure}
\begin{center}
\mbox{\epsfig{figure=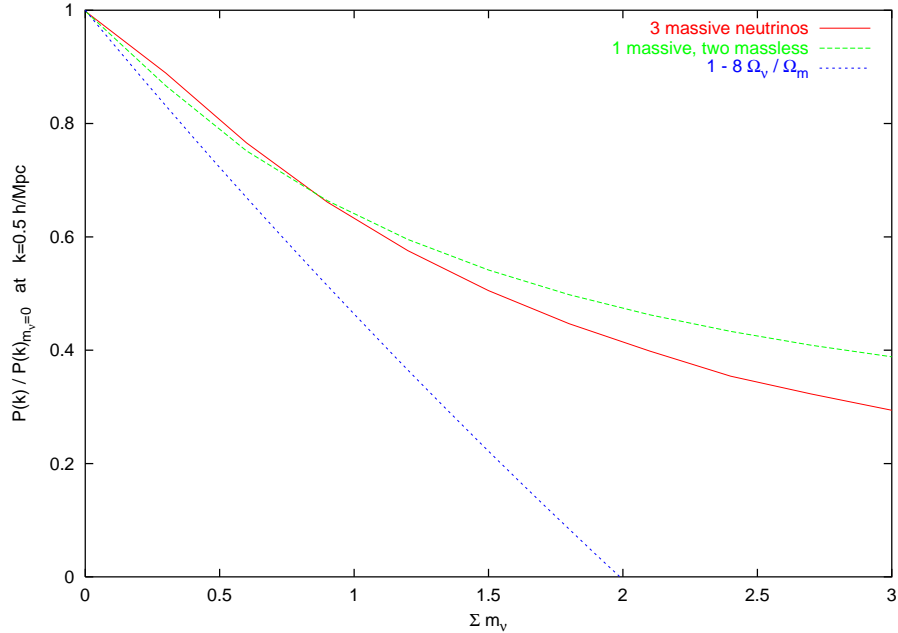,angle=-90,width=0.8\textwidth}}
\end{center}
\caption{Ratio between the power spectrum at $k=0.5 h$ Mpc$^{-1}$ with
and without massive neutrinos, as a function of the total mass $\sum
m_\nu$ in eV when the number of massive neutrinos is either three or
one. The other cosmological parameters are kept fixed and the universe
is assumed to be flat. The analytical approximation often quoted in
the literature is also shown (blue dashed).
\label{fig:free-stream}}
\end{figure}

Nowadays we know that neutrinos can not constitute the main DM
component, and the main question now is to find how large the neutrino
contribution (or HDM) can be.  For values of the neutrino masses of
order eV, the free-streaming effect can be detectable in the linear
power spectrum of LSS, reconstructed from redshift survey experiments.
This effect is shown in Fig.\ \ref{fig:free-stream}, where the
suppression of the mass power spectrum $P(k)$ caused by massive
neutrinos was calculated for two cases with the same $\sum m_i$, and
compared with the analytical approximation \cite{Hu:1997mj} $\Delta
P(k)/P(k) \sim -8 \Omega_\nu/\Omega_m$, usually quoted in the
literature. Thus cosmology offers a unique opportunity to measure the
absolute value of the neutrino masses.

The free-streaming effect is more important than the direct effect of
the contribution of massive neutrinos to the energy density of the
universe: different values of the neutrino density fraction
$\Omega_{\nu}$ have to be compensated by different values of the other
components (cold dark matter, baryons, dark energy or eventually the
curvature parameter).  These changes in the cosmic balance can modify
some characteristic times and scales in the history of the universe,
like the time of equality between matter and radiation, or the size of
the Hubble radius at photon decoupling. Neutrino mass, however,
influences only slightly the spectrum of CMB
anisotropies. Nevertheless, it is crucial to combine CMB and LSS
observations in order to measure the neutrino mass, because CMB data
give independent constraints on the cosmological parameters, and
partially removes the parameter degeneracy that would arise in an
analysis of the matter power spectrum only.

These effects of neutrino masses probed by CMB and LSS depend, in
first approximation, on the sum of the masses $\sum m_{\nu}$ if all
neutrino states have the same number density, that of
Eq.~(\ref{nunumber}). Before WMAP data, the best bound on $\sum
m_{\nu}$ was obtained using the power spectrum of galaxy clustering
from the 2dFGRS and adding other independent cosmological constraints
through priors on the other parameters \cite{Elgaroy:2002bi}. The
upper limit found was $\sum m_{\nu}< 2.2$ eV (95\% CL).

At present, the most stringent upper limit on the sum of neutrino
masses was reported by the WMAP collaboration \cite{Spergel:2003cb},
using the first-year WMAP data, CMB data from other experiments (ACBAR
and CBI), the LSS data from 2dFGRS and the matter power spectrum on
small scales inferred from the Lyman $\alpha$ forest. The limit is
$\sum m_{\nu}< 0.7$ eV (95\% CL), approximately a factor three better
than the previous 2dFGRS bound. 

Two other recent analyses \cite{Hannestad:2003xv,Elgaroy:2003yh} have
calculated the upper bound on $\sum m_{\nu}$ within a more
conservative approach. In particular the Lyman $\alpha$ data was not
used, since it has been recently pointed out that the errors were
probably underestimated \cite{Seljak:2003jg}. Hannestad also used a
completely free bias parameter (which relates the matter and the
galaxy correlation spectra), emphasizing the importance of the HST
prior on $h$ in constraining the neutrino mass. The upper bound
found \cite{Hannestad:2003xv} is
\begin{equation}
\sum m_{\nu}< 1.0~{\rm eV~(95\%~CL)}
\label{mnuconservative}
\end{equation}
which should be considered conservative, but still dependent on the
set of cosmological parameters (a minimal set) and the assumed priors
(see the discussion on \cite{Hannestad:2003xv,Elgaroy:2003yh}).

It should be noted that this cosmological bound on neutrino masses is
comparable and even stronger than direct terrestrial limits, presently
at the eV order. A direct experimental limit on the neutrino mass
comes from the experiments searching for kinematical effects on the
spectrum of tritium $\beta$ decay \cite{weinheimer} ($m_\nu<2.2$ eV at
$95 \%$ CL), a bound expected to be improved by the KATRIN experiment
to reach $0.35$ eV. A complementary bound can be put on the neutrino
masses if they are Majorana particles, from the non-observation
\footnote{It has been claimed a positive evidence of neutrinoless
double beta decay, criticized by other authors. For a discussion, see
e.g.\ \cite{strumia}.} of neutrinoless double beta decay
\cite{petcov}.

The cosmological effect of neutrino masses is important because
presently we have experimental evidences of flavour neutrino
oscillations, which come from solar, atmospheric and reactor neutrino
experiments (see e.g.\ \cite{fogli,suzuki}). They are sensitive to mass
differences between the three neutrino mass eigenstates
$(m_1,m_2,m_3)$.  Approximate $3\sigma$ ranges for the squared mass
differences are $\Delta m_{\rm atm}^2=1.2-4.8\times 10^{-3}$ eV$^2$
and $\Delta m_{\rm sun}^2=5.1-19\times 10^{-5}$ eV$^2$, respectively
\cite{Pakvasa:2003zv}.

These values are perfectly compatible with a hierarchical scenario
where the lightest neutrino state has $m_1\sim 0$, $m_2\sim (\Delta
m_{\rm sun}^2)^{1/2}$ and $m_3\sim (\Delta m_{\rm atm}^2)^{1/2}$ (or
with an inverted hierarchy where $m_3 \sim m_2 \sim (\Delta m_{\rm
atm}^2)^{1/2}$, separated by the small $\Delta m_{\rm sun}^2$).  The
largest neutrino mass would then be of the order $(\Delta m_{\rm
atm}^2)^{1/2}\sim 0.05$ eV. However, it is still possible that the
three neutrino states are degenerate, with masses much larger than the
differences. In such a case, $\sum_i m_i = m_1+m_2+m_3 \simeq 3m_0$.
Cosmology is the most powerful tool for constraining the mass scale
$m_0$ in the degenerate scenario. Applying the result from
Eq.~(\ref{mnuconservative}), we get $m_0 \leq 0.33$ eV.

At the current experimental level it is yet possible to reach the
needed precision to test the non-degenerate neutrino schemes described
above. However, future CMB observations by WMAP and PLANCK, among
others, combined with LSS data from larger galaxy surveys will enhance
the cosmological sensitivity to neutrino mass. The pioneering
calculation in ref.\ \cite{Hu:1997mj} found that the combination of
future CMB and LSS data, and in particular the Sloan Digital Sky
Survey (SDSS) will push the bound on $\sum m_\nu$ to approximately
$0.3$ eV.

More recently, Hannestad \cite{Hannestad:2002cn} has updated the
forecast analysis and concluded that the PLANCK CMB experiment
combined with the SDSS data could measure a neutrino mass of $0.12$ eV
at $95\%$ CL, almost reaching the values in the hierarchical
scenarios. With an order of magnitude larger survey volume than SDSS,
the limit could be pushed down to $0.03-0.05$ eV.

There exist other cosmological probes of neutrino masses that could,
in the next future, reach similar sensitivities. One can use the weak
gravitational lensing of background galaxies by intervening matter to
probe the mass distribution of the universe. It has been estimated
\cite{Hu:2002rm,Abazajian:2002ck} that a $0.1$ eV sensitivity on
$m_\nu$ could be reached, depending on the equation of state of dark
energy. Alternatively, the authors of ref.\ \cite{Kaplinghat:2003bh}
describe how to use the distortions of CMB temperature and
polarization spectra caused by gravitation lensing to get a limit on
$m_\nu$. Their calculations show that again a $\sim 0.1$ eV value
could be reach with PLANCK data.

The cosmological limit on neutrino masses also applies for neutrino
mass schemes that explain the results of LSND, a short-baseline
accelerator experiment which gives an excess of $\bar{\nu}_e$ events.
These results constitute an additional evidence of flavour neutrino
conversions than can be explained if $\Delta m^2_{\rm LSND} = 0.2-7$
eV$^2$.  The minimal way to accommodate for LSND is to include a
fourth sterile neutrino, and it was shown
\cite{strumia,Maltoni:2002xd} that only the so-called 3+1 scheme,
where one heavy neutrino state of mass $m_4 \simeq (\Delta m^2_{\rm
LSND})^{1/2}\simeq \sum m_\nu$ exists, in addition to the three
flavour neutrinos, is still marginally allowed.  The LSND results
would be checked by the ongoing MiniBoone experiment \cite{louis}.

We have discussed before that the four neutrino scheme is disfavoured
by BBN. In addition, shortly after the WMAP results
\cite{Spergel:2003cb}, several papers (see e.g.\
\cite{Pierce:2003uh,Giunti:2003cf}) emphasized that the bound $\sum
m_{\nu}< 0.7$ eV was in conflict with the allowed windows for $m_4$ in
a 3+1 scheme.  We have discussed that this limit is probably too
restrictive. It turns out that, in any case, this value or that in
Eq.~(\ref{mnuconservative}) can not be applied in the LSND case
because it was calculated assuming three neutrinos. A calculation of
the upper bound on $\sum m_{\nu}$ including four neutrino states was
done in ref.\ \cite{Hannestad:2003xv}. It was found that the effect of
an extra neutrino can partially compensate the effect of the neutrino
mass, leading to a bound $\sum m_{\nu}< 1.38$ eV (95\% CL, $N_{\rm
eff}=4$). This value does not exclude yet the LSND windows with the
lower $m_4$ values.
  
\section{Conclusions}  

The existence of a relic neutrino sea in the universe allows us to
study bounds on neutrino properties that are complementary to
terrestrial experiments. We have discussed two different bounds that
have been improved after the recent WMAP data, namely those on the
effective number of relativistic neutrino species and on the sum of
neutrino masses. Both are nice examples of the connection between
cosmology and particle physics. The bound on $\Delta N_{\rm eff}$
restricts the existence of new relativistic particles or new decay
processes beyond the Standard Model. The limit on $\sum m_\nu$ fixes
the absolute neutrino mass scale, to which the current experimental
evidences on flavour neutrino oscillations are not sensitive.

\section{Acknowledgements}
The author was supported by the Spanish grant BFM2002-00345, the ESF
Network {\em Neutrino Astrophysics} and a Marie Curie fellowship under
contract HPMFCT-2002-01831.

\end{document}